\documentclass[twocolumn]{aastex631}
\usepackage{amsmath,amssymb}
\usepackage{booktabs}
\usepackage{graphicx}
\usepackage{hyperref}
\shorttitle{Single-epoch Low-rank RFI Non-identifiability}
\shortauthors{Kim}

\begin{document}

\title{Operational Non-identifiability of Single-epoch Low-rank RFI Mitigation:\\
Controlled Failure-mode Analysis and HERA Evidence}

\author[0009-0006-9745-2960]{Yujin Kim}
\affiliation{Department of Data Science, Jeju National University, Jeju, South Korea}
\correspondingauthor{Yujin Kim}
\email{europa5218@stu.jejunu.ac.kr}

\begin{abstract}
Low-rank decomposition (singular value decomposition / principal component analysis based) is widely used as a first-line tool for mitigating
radio-frequency interference (RFI) in low-frequency radio astronomy, both in operational
pipelines and methodological studies.
In the \emph{single-epoch} regime---when only one time--frequency dynamic spectrum
$D(t,\nu)$ is available---there is no structural guarantee that science and interference
can be separated by rank selection alone.
We model the observation as $D=S+I+N$, where $S$ is approximately low-rank science signal,
$I$ is sparse but spectrally structured RFI, and $N$ is noise.
When $S$ and $I$ substantially share the same singular-vector subspace (mixed singular directions),
no choice of rank can simultaneously suppress residual RFI contamination and preserve the science signal; we
call this \emph{operational non-identifiability}.
Using controlled synthetic experiments with satellite-like harmonic-comb RFI and weak spectral
features, and 420 calibrated HERA snapshots over 50--225\,MHz, we show that truncated SVD can
produce band-wide bias, while frequency-weighted SVD shifts the trade-off to safer operating
points but does not remove the bias floor.
Rather than proposing a new subtraction method, we provide compact diagnostics---mixed-mode
inspection, rank sweeps, and Pareto views---to assess single-epoch low-rank cleaning in
real-world pipelines.
\end{abstract}

\keywords{radio astronomy --- radio frequency interference --- reionization --- data analysis}

\section{Introduction}\label{sec:intro}

Low-frequency radio observations increasingly face spectrally structured radio-frequency interference (RFI), including unintentional electromagnetic radiation (UEMR) from low-Earth-orbit (LEO) satellite constellations, which can contaminate Epoch of Reionization (EoR) experiments \citep[e.g.,][]{FridmanBaan2001, DiVruno2023a, Bassa2024}.
For arrays such as HERA \citep{DeBoer2017, Abdurashidova2022}, the cosmological 21\,cm signal is fragile to weak contamination and subject to stringent protection criteria \citep[e.g.,][]{ITUR2005RA769}.
Because of their speed and interpretability, low-rank subspace methods based on singular value decomposition (SVD) and principal component analysis (PCA) are commonly used in pipelines, often alongside automated flagging (e.g., AOFlagger) \citep{Offringa2010, Offringa2012} and related mitigation techniques \citep{Leshem2000, FridmanBaan2001}.
Deep-learning approaches can perform RFI detection/mitigation on time--frequency data
\citep[e.g.,][]{Akeret2017, Wilensky2019, Kerrigan2019, Yang2020, VafaeiSadr2020, Connor2018},
but typically rely on large labeled data sets and diversity across epochs or instruments.
This work instead targets the \emph{single-epoch}, fixed-baseline configuration.

\subsection{Relation to prior work}\label{subsec:relation}

A substantial literature uses low-rank and low-rank$+$sparse decompositions for mitigating
foregrounds and RFI \citep[]{Leshem2000, FridmanBaan2001, Candes2011, Chandrasekaran2011},
and related extensions such as independent component analysis (ICA), non-negative matrix factorization (NMF), and robust principal component analysis (RPCA) variants for 21\,cm analysis and RFI
suppression \citep[e.g.,][]{Datta2010, Parsons2012, Morales2012, Pober2014s}.
Many formulations implicitly rely on conditions that limit subspace overlap between the
science and contaminant components and on perturbation bounds for singular vectors
\citep[e.g.,][]{Wedin1972, GolubVanLoan2013}.
This paper does not propose a new algorithm.
Instead, we ask a meta-level question:
in the single-epoch regime---a single dynamic spectrum for a fixed baseline and polarization---are
the science and interference components operationally identifiable under any choice of rank?
We show that when the science signal and structured RFI share singular-vector directions,
rank selection cannot simultaneously achieve low residual contamination and low distortion.
This limitation is therefore not specific to one implementation, but reflects a structural constraint
on single-epoch, subspace-based cleaning.

\subsection{Single-epoch focus and contributions}\label{subsec:single_epoch}

\begin{figure*}
  \centering
  \includegraphics[width=\textwidth]{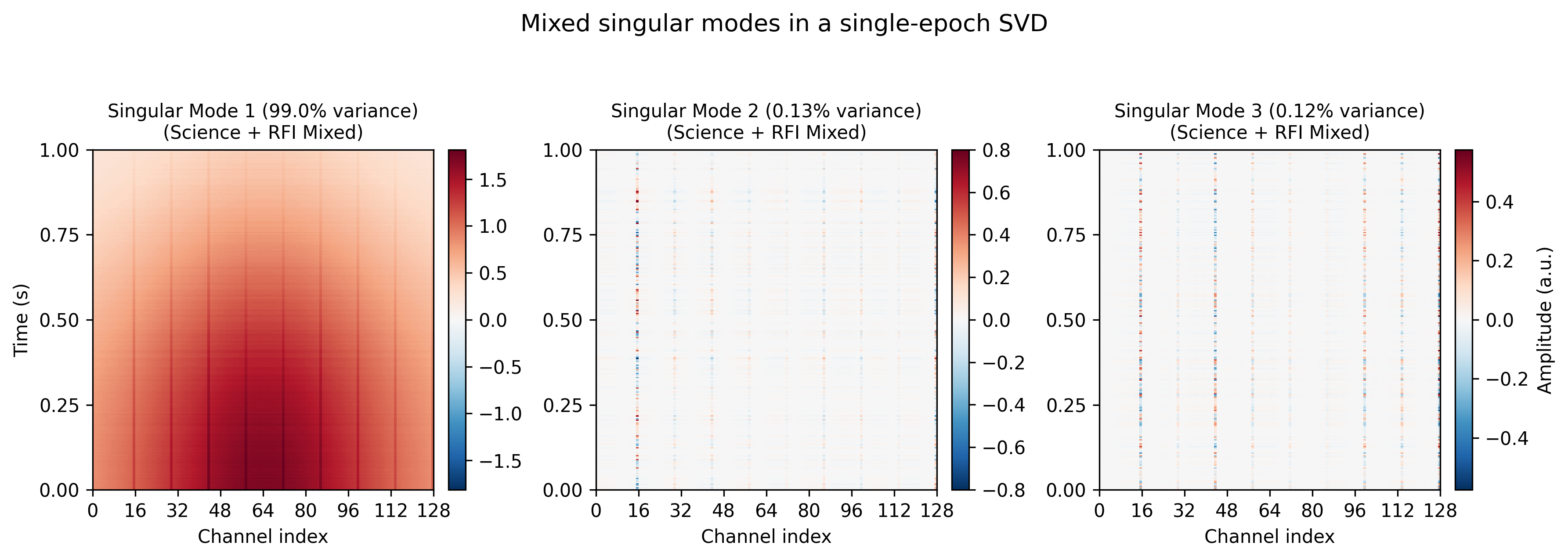}
  \caption{Synthetic illustration of the failure mechanism of single-epoch low-rank cleaning.
    (a) A toy single-epoch time--frequency matrix $X$ composed of a smooth foreground-like component, intermittent comb-like narrowband RFI, a faint structured science ripple, and noise.
    (b) First singular-mode reconstruction, $X_1 = \sigma_1 u_1 v_1^\top$, capturing the dominant mixed structure.
    (c) Second singular-mode reconstruction, $X_2 = \sigma_2 u_2 v_2^\top$, illustrating additional mixed components.
    For visual clarity, this figure uses a visualization-oriented realization based on the same components described in Table~\ref{tab:synthetic-params} and Section~\ref{sec:band-metrics}, with simplified amplitudes and enhanced contrast; it is intended for qualitative interpretation rather than quantitative evaluation.
    Leading singular modes do not correspond to physically separable components, but instead span mixed directions that simultaneously encode astrophysical signal and RFI, illustrating the operational non-identifiability of single-epoch low-rank separation.
  }
  \label{fig:singular_mode_mixing}
\end{figure*}

We focus on decisions made from a single time--frequency matrix, without time/baseline diversity
or strong external priors (e.g., ephemerides; cf.~\citealt{Dillon2020, Abdurashidova2022}).
In this setting, subspace overlap can dominate performance.
Using controlled synthetic experiments and HERA snapshots, we expose operational
non-identifiability via mixed-mode inspection, rank sweeps, and Pareto diagnostics, and we
propose a minimal, reproducible quality-assurance (QA) framework for reporting the operational risk of any chosen
single-epoch cleaning configuration.
Figure~\ref{fig:singular_mode_mixing} provides a synthetic illustration of the single-epoch failure mechanism considered in this paper: when foreground-like and RFI-like structure co-occupy the leading singular mode, low-rank subtraction cannot cleanly separate contamination removal from scientific preservation.

\section{Problem setup}
\label{sec:setup}

\subsection{Single-epoch observation matrix}

We consider a single-epoch time--frequency snapshot obtained by fixing an
interferometric baseline and polarization.
The observation is represented as a matrix $D \in \mathbb{R}^{T \times F}$
with $T$ time samples and $F$ frequency channels,
\begin{equation}
  D(t,\nu) = S(t,\nu) + I(t,\nu) + N(t,\nu),
\end{equation}
where $S(t,\nu)$ is the scientific component, $I(t,\nu)$ is structured RFI,
and $N(t,\nu)$ is noise.
In the single-epoch regime we assume no additional diversity from repeated
epochs or external information (e.g., satellite ephemerides or auxiliary
observations): $D$ alone must support any separation between $S$ and $I$.

\subsection{Low-rank (subspace) cleaning operator}
\label{sec:lowrank_operator}

Let $D = U \Sigma V^\top$ be the SVD, with
left singular vectors $U \in \mathbb{R}^{T\times T}$, right singular vectors
$V \in \mathbb{R}^{F\times F}$, and singular values
$\Sigma = \mathrm{diag}(\sigma_1,\ldots,\sigma_{\min(T,F)})$ ordered
non-increasingly.
For a chosen rank $k$, the truncated SVD approximation is
\begin{equation}
  D_k = U_{1:k} \Sigma_{1:k} V_{1:k}^\top ,
\end{equation}
and the corresponding residual (``cleaned'' result) is
\begin{equation}
  D_{\mathrm{clean}}(k) = D - D_k .
\end{equation}
Throughout the paper we use ``truncated SVD'' and ``standard SVD cleaning''
interchangeably to denote this rank-$k$ truncation of the SVD
\citep[e.g.,][]{GolubVanLoan2013}. We avoid the abbreviation ``TSVD'' to prevent confusion with the temporally smoothed variant (TempSVD) introduced below.

In the idealized low-rank\,+\,sparse setting studied in the robust PCA literature, the scientific
component $S$ and the interference component $I$ are assumed to occupy sufficiently disjoint or
\emph{incoherent} subspaces, and $I$ is assumed to be sparse in the canonical time--frequency basis
(e.g., \citealt{Candes2011,Chandrasekaran2011}). Under these assumptions, there exists a decomposition
in which the low-rank term captures $S$ and the sparse term captures $I$, and, equivalently, there
exists a rank choice $k_\star$ for which $D_{\mathrm{clean}}(k_\star)$ is science-dominated while
$D_{k_\star}$ is RFI-dominated.

In practice, a variety of data-driven rules have been proposed for choosing $k$, including
profile-likelihood and scree-plot methods \citep[e.g.,][]{Zhu2006GhodsiDimensionalitySelection},
optimal hard thresholds on singular values \citep[e.g.,][]{Gavish2014OptimalHardThreshold},
and eigenvalue-distribution criteria for factor models
\citep[e.g.,][]{Onatski2010DeterminingNumberFactors}. These approaches are extremely useful
when the underlying low-rank structure is well separated from noise or when the goal is
dimension reduction, but they implicitly rely on assumptions that do not hold in the regime
we study here---most notably, that the dominant singular subspace can be interpreted as either
``signal'' or ``interference'' but not as a mixture of both.

Our focus in this work is the opposite regime: \emph{mixed singular directions},
where $S$ and $I$ substantially overlap in the singular-vector subspaces
$U$ and $V$. In this mixed-subspace regime, each dominant singular mode contains a
nontrivial combination of scientific signal and structured RFI.
As a result, varying the truncation rank $k$ inevitably trades off
science preservation against interference suppression.
Accordingly, we interpret the rank $k$ not as a uniquely determined
quantity, but as an operational hyperparameter whose effect must be
assessed empirically in the single-epoch setting.
We therefore treat $k$ as an \emph{operational} hyperparameter that moves configurations
along a Pareto front between under-cleaning and over-cleaning, rather than as a quantity
that can be uniquely determined from single-epoch statistics alone.

\section{Band Definition and Evaluation Metrics}
\label{sec:band-metrics}

\subsection{Fixed Science Band and Core}
\label{sec:fixed-band-core}

To avoid ``moving goalposts,'' where evaluation windows shift between
experiments, we explicitly fix the science band and protected core
throughout all synthetic tests.
In the controlled toy experiments, we use $F = 240$ frequency channels
spanning $0$--$12~\mathrm{MHz}$, with a channel spacing of
\begin{equation}
\Delta \nu = 0.05~\mathrm{MHz}.
\end{equation}
This spacing was chosen so that each synthetic comb-RFI line
($\sigma_\nu = 0.02$\,MHz; Table~\ref{tab:synthetic-params}) is sampled by at
least two channel widths, a minimal Nyquist-like criterion.
The comb line width of 20\,kHz is comparable to the narrowband
unintentional electromagnetic radiation (UEMR) features reported in
satellite observations, where individual spectral peaks are confined
within single 12\,kHz channels
\citep[e.g.,][]{DiVruno2023a, Bassa2024}.
We define three disjoint spectral regions:
\begin{itemize}
\item \textbf{Science band (evaluation window)}: 5.5--6.5 MHz.
\item \textbf{Science core (protected region)}: 5.8--6.2 MHz.
\item \textbf{Outside band (fitting region)}: all remaining channels.
\end{itemize}
Conceptually, the science band is the frequency window over which
recovery performance is evaluated, while the science core represents
the highest-priority region whose spectral fidelity is most critical.
The outside band is primarily used to learn low-rank structure and
interference patterns.
The fixed science band and protected core play a role analogous to
protected EoR windows used in foreground-avoidance analyses
(e.g., \citealt{Datta2010,Parsons2012,Morales2012,Pober2014s}),
but are defined here purely for controlled single-epoch diagnostics
rather than for direct cosmological interpretation.

\subsection{Primary Metric: Relative Bias}
\label{sec:relative-bias}

Our primary metric is the channel-wise relative bias,
defined as
\begin{equation}
B(\nu) = 100 \times
\frac{\hat{S}(\nu) - S(\nu)}{\max\bigl(|S(\nu)|,\epsilon\bigr)},
\label{eq:relative-bias}
\end{equation}
expressed in percent.
Here $\hat{S}(\nu)$ is the recovered spectrum obtained from the
rank-$k$ cleaned dynamic spectrum $D_{\mathrm{clean}}^{(k)}$,
$S(\nu)$ is the injected (true) science spectrum, and
the stabilization constant $\epsilon = 10^{-5}$ is a small denominator floor introduced to avoid
unstable relative-bias values in channels where the injected science
spectrum is close to zero. In our normalized setup, this floor is used
only to regularize near-zero denominators and does not set the scale of
the reported bias in channels where $|S(\nu)| \gg \epsilon$.
The recovered spectrum $\hat{S}(\nu)$ is obtained by time-averaging the residual matrix
$D_{\mathrm{clean}}^{(k)}$ along the time axis using a NaN-safe mean when masking is applied.
We also track the absolute error
\begin{equation}
A(\nu) = \hat{S}(\nu) - S(\nu),
\label{eq:absolute-error}
\end{equation}
which complements the relative bias by retaining information about
the absolute scale of the distortion.
Both $B(\nu)$ and $A(\nu)$ are summarized over the science band and
science core using medians, percentiles, or simple channel-wise sums,
depending on the diagnostic being performed.

\subsection{Auxiliary Metrics}
\label{sec:aux-metrics}

To connect Monte Carlo summaries to scientific impact, we use two
additional aggregate metrics.
First, we define the integrated absolute error over the science core as
\begin{equation}
\mathrm{IAE}_{\mathrm{core}} =
\sum_{\nu \in \mathrm{core}}
\bigl|\hat{S}(\nu) - S(\nu)\bigr|,
\label{eq:iae-core}
\end{equation}
which serves as a channel-summed proxy (i.e., a discrete sum over channels, with no explicit $\Delta\nu$ factor)
within the protected region.

Second, we define the detection failure rate as
\begin{equation}
p_{\mathrm{fail}} = \frac{N_{\mathrm{fail}}}{N_{\mathrm{trial}}},
\label{eq:pfail}
\end{equation}
where $N_{\mathrm{trial}}$ is the total number of injection--recovery trials and
$N_{\mathrm{fail}}$ is the number of trials in which a Gaussian line fit to the recovered
spectrum within the science core either fails to converge or yields a fitted amplitude
below the adopted detection threshold.
This metric quantifies missed recovery of the injected science feature rather than generic spectral-shape errors.

\section{Methods}
\label{sec:methods}

We now define the low-rank cleaning methods considered for a single-epoch matrix
$D \in \mathbb{R}^{T \times F}$ and the comparison references.

\subsection{Frequency-weighted SVD (FWSVD)}
\label{sec:rank-sweep}

To suppress variance explanation in the protected channels, we introduce a
diagonal weight matrix $W \in \mathbb{R}^{F \times F}$ along the frequency
axis. Channel weights are
\begin{equation}
  w(\nu) =
  \begin{cases}
    w_{\mathrm{core}}, & \nu \in \mathrm{science~core}, \\
    w_{\mathrm{prot}}, & \nu \in \mathrm{science~band}\setminus\mathrm{core}, \\
    1,                 & \text{otherwise},
  \end{cases}
\end{equation}
with $0 < w_{\mathrm{core}} \leq w_{\mathrm{prot}} \leq 1$, and
$W = \mathrm{diag}\bigl(w(\nu)\bigr)$.
We define the frequency-weighted snapshot as
\begin{equation}
D^{(W)} \equiv D\,W,
\label{eq:D_weighted}
\end{equation}
and compute its SVD,
\begin{equation}
  D^{(W)} \approx \tilde{U} \tilde{\Sigma} \tilde{V}^\top .
\end{equation}
For rank $k$, the rank-$k$ approximation in the weighted space is
$\tilde{U}_{1:k} \tilde{\Sigma}_{1:k} \tilde{V}_{1:k}^\top$, and we subtract
\begin{equation}
  D_{\mathrm{clean}}^{\mathrm{(FWSVD)}}(k)
    = D - \tilde{U}_{1:k} \tilde{\Sigma}_{1:k} \tilde{V}_{1:k}^\top W^{-1}.
  \label{eq:dclean_fws}
\end{equation}

\subsection{Temporally smoothed SVD (TempSVD)}

To benchmark the effect of imposing temporal regularity on the fitted low-rank component,
we include a simple temporally smoothed SVD reference method.
Let $K\in\mathbb{R}^{T\times T}$ be a linear smoothing operator acting along time (we use a symmetric
boxcar moving-average window of width $m$ samples, normalized to unit row sum).
We form the smoothed matrix
\begin{equation}
  D^{(\mathrm{sm})} = K D,
\end{equation}
compute its rank-$k$ truncated-SVD approximation $(D^{(\mathrm{sm})})_k$, and subtract this fit from the
original (unsmoothed) snapshot,
\begin{equation}
  D_{\mathrm{clean}}^{(\mathrm{TempSVD})}(k) = D - (D^{(\mathrm{sm})})_k .
  \label{eq:dclean_tempsvd}
\end{equation}
The window width $m$ controls the strength of temporal averaging: larger $m$ suppresses fast time variability
but can introduce horizontal (time-smeared) residual structure. We treat TempSVD as an illustrative
regularized reference method, not as a proposed default.

In practice, for TempSVD we use an alternating least-squares solver for the rank-1 case for numerical stability and computational efficiency, while retaining the same conceptual structure.

To summarize the trade-off between science distortion and residual RFI contamination, we define two scalar Pareto proxies over the science band.
For clarity, `residual RFI contamination' denotes signal remaining after cleaning, not instrumental spectral leakage or ADC artifacts.

The science-distortion proxy,
\begin{equation}
  \mathrm{Bias}^{-}
  =
  1 -
  \frac{\sum_{\nu \in \mathrm{band}} \hat{S}(\nu)}{\sum_{\nu \in \mathrm{band}} S(\nu)} ,
\end{equation}
measures the signed fractional change of the injected science component after cleaning
($\mathrm{Bias}^{-}=0$ for perfect preservation). Negative values correspond to net amplification rather than loss.
In Pareto plots we therefore use $|\mathrm{Bias}^{-}|$ as a distortion magnitude, while retaining the sign in tabulated summaries.

The residual-contamination proxy,
\begin{equation}
  \mathrm{Bias}^{+}
  =
  \frac{\sum_{\nu \in \mathrm{band}} \hat{I}(\nu)}{\sum_{\nu \in \mathrm{band}} I(\nu)} ,
\end{equation}
is evaluated using dedicated RFI-only trials in which the science component is set to $S=0$,
so that $D = I + N$. In these trials, the time-averaged residual
$\hat{I}(\nu) = \bigl\langle D_{\mathrm{clean}}^{(k)}(t,\nu) \bigr\rangle_t$
directly estimates the residual RFI contamination that survives rank-$k$ subtraction.
The denominator $\sum_{\nu \in \mathrm{band}} I(\nu)$ is the corresponding time-average of the
injected RFI spectrum $I(t,\nu)$, which is known in the synthetic testbed.
Smaller values of $\mathrm{Bias}^{+}$ and $\mathrm{Bias}^{-}$ are desirable;
in practice, mixed subspaces prevent the Pareto front from approaching $(0,0)$.

\subsection{Additional reference methods and masking}

To test whether the observed behavior is SVD-specific, we also include
NMF \citep{LeeSeung1999},
ICA \citep{Hyvarinen2000}, and
RPCA \citep{Candes2011}. In the overlap regimes considered
here, NMF frequently fails to converge stably, and RPCA often exceeds
iteration and time budgets; moreover, the incoherence and sparsity assumptions
underpinning RPCA are explicitly violated when science and RFI share the
same singular subspace. We therefore report core-bias statistics only for
truncated SVD, FWSVD, ICA, and a conservative hard-mask reference, and treat NMF/RPCA
as negative results that support the same non-identifiability picture.
As a conservative reference, hard masking sets contaminated
time--frequency pixels to NaN and estimates the spectrum via NaN-robust
averages. This sacrifices coverage to prioritize scientific integrity and
serves as an external comparison for residual-contamination control rather than a point on the low-rank
Pareto front.

\subsection{Operational QA recipe}

We propose the following minimal QA recipe:
\begin{itemize}
  \item Fix and document the science band and core.
  \item For any configuration, report (i) a science-core bias metric
  (e.g., median $B(\nu)$ or $\mathrm{IAE}_{\mathrm{core}}$; see Equations~\eqref{eq:relative-bias} and~\eqref{eq:iae-core}) and
  (ii) a detection failure rate $p_{\mathrm{fail}}$ (Equation~\eqref{eq:pfail}) across realizations or snapshots.
  \item When masking is used, also report the retained coverage in the science band and core.
\end{itemize}
Indicative thresholds tuned on our synthetic testbed are median core bias
$\lesssim 20\%$ and $p_{\mathrm{fail}}\lesssim 5\%$; if these are not met,
we recommend falling back to conservative masking and, where possible,
deferring interpretation to multi-epoch or multi-baseline analyses.

\section{Results on synthetic experiments}
\label{sec:synthetic_results}

We now present results from the synthetic single-epoch testbed, focusing on
(i) mixed singular modes, (ii) reference method behavior under controlled
overlap, and (iii) rank and weight sensitivity summarized through Pareto
geometry. The configuration is summarized in Appendix~\ref{app:synthetic}.

\subsection{Reference method comparison under controlled overlap}

For a representative overlap configuration (e.g., ${\rm SNR}=1.0$ with
high-dynamic-range comb amplitude), we compare methods using relative
bias on the science core.
A typical Monte Carlo summary ($N_{\mathrm{trial}}=200$ trials) is shown in
Table~\ref{tab:bias_comparison}.

\begin{table}[t]
\centering
\caption{Relative bias on the science core ($N_{\mathrm{trial}} = 200$).}
\label{tab:bias_comparison}
\begin{tabular}{lrrrr}
\toprule
Method   & Median & Mean   & Max      & Trials \\
\midrule
Truncated SVD & 70.83  & 109.28 & 572.12   & 200 \\
FWSVD          & 62.63  &  93.52 & 571.90   & 200 \\
ICA            & $5.8\times10^4$ & $2.2\times10^5$ & $9.8\times10^5$ & 200 \\
Hard mask      & N/A    & N/A    & N/A      & 200 \\
\bottomrule
\end{tabular}
\end{table}

FWSVD reduces the median and mean bias relative to standard truncated SVD and suppresses
catastrophic failures, but substantial distortions persist in the overlapping
regime. Even in the best case, the median science-core bias remains at
$\simeq 60\%$, reinforcing the presence of a non-zero bias floor.
In contrast, the ICA reference exhibits a qualitatively different, pathological
failure mode: the median and mean science-core biases blow up to
$\sim 10^{4}$--$10^{5}\,\%$, and the maximum bias reaches
$\sim 10^{6}\,\%$ (Table~\ref{tab:bias_comparison}).
This behavior is consistent with known limitations of ICA in extremely
high-dynamic-range mixtures.
The algorithm assumes that the latent components are statistically independent
and non-Gaussian, and it is only defined up to arbitrary rescaling and
permutation of the recovered sources.
In our controlled single-epoch setting, a very bright, spectrally structured
RFI component, Gaussian thermal noise, and a faint, approximately Gaussian
cosmological signal are linearly mixed with dynamic range $\gtrsim 10^{4}$.
Under these conditions, the scale indeterminacy of ICA and the contrast
function used to enforce non-Gaussianity can drive the optimization into
local minima in which the faint science component is effectively absorbed or
over-subtracted in one of the recovered independent components.
The resulting demixing matrix amplifies small modeling errors by orders of
magnitude, producing the catastrophic bias values reported in the table.
NMF and RPCA exhibit similar sensitivity to the extreme dynamic range and
masking, frequently suffering from convergence and runtime failures, and we
therefore treat them only as qualitative supporting evidence for
non-identifiability rather than as operational reference methods.

\subsection{Rank sensitivity}

\begin{figure}[t]
  \centering
  \includegraphics[width=\columnwidth]{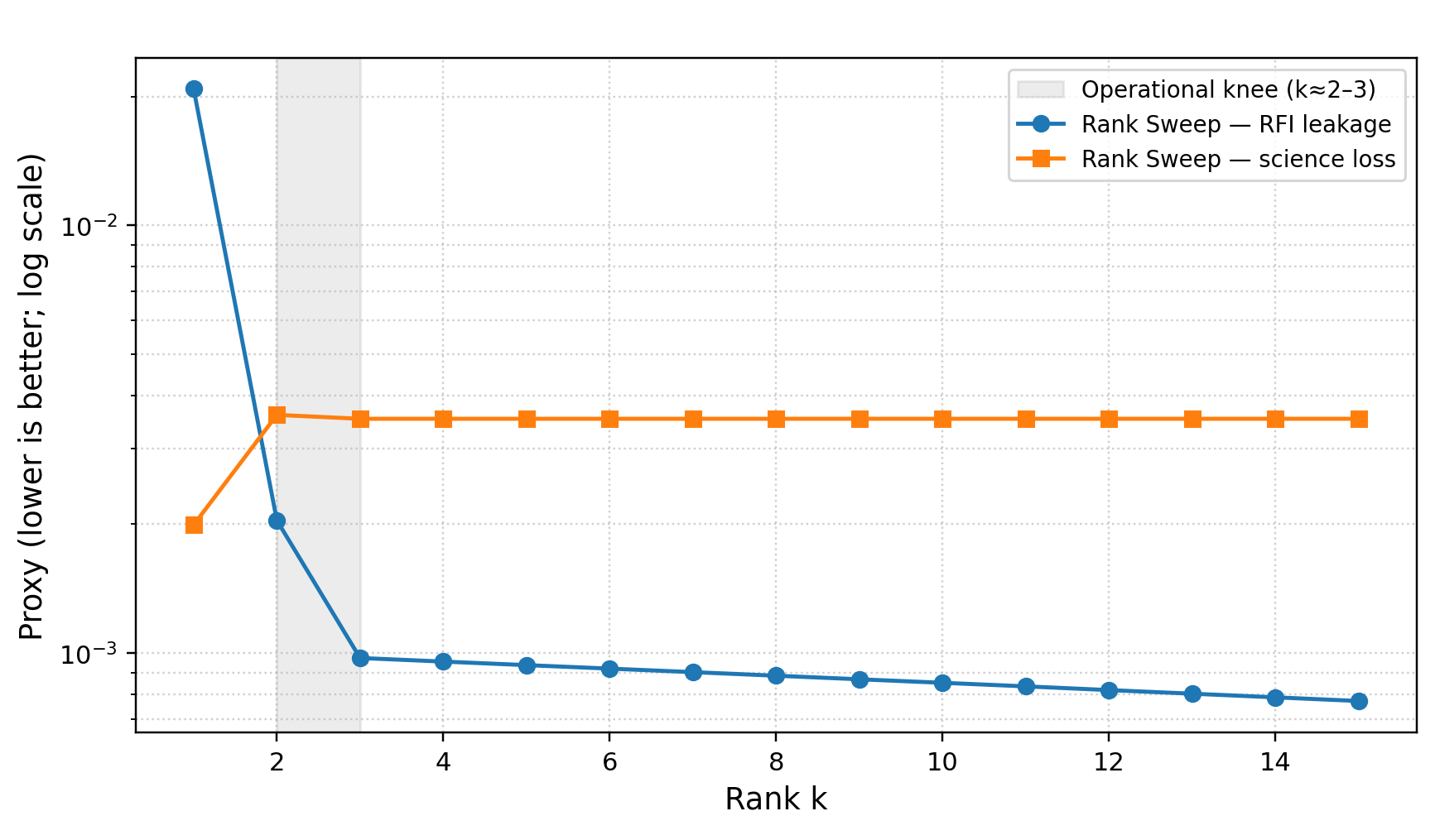}
  \caption{\textbf{Rank sweep on the synthetic comb + broadband ripple + sloped background case.}
        The trade-off between RFI leakage and science distortion is shown as a function of rank $k$.
        Lower ranks suppress large-scale structure but retain RFI, while higher ranks increasingly distort the science signal.
  }
  \label{fig:rank_synth}
\end{figure}

Rank sweeps reveal a monotonic trade-off between residual RFI contamination and science loss:
increasing rank suppresses interference while increasingly distorting the
science signal (Figure~\ref{fig:rank_synth}).
The same trade-off is shown in Pareto projection in
Figure~\ref{fig:bias_tradeoff_panels}a, where each point represents a
different rank and the two axes correspond to the contamination and distortion
proxies from Figure~\ref{fig:rank_synth}.
No extended rank range exists in which both effects are simultaneously small,
so rank choice becomes a risk-management decision rather than a pure optimization.

\subsection{Weight sensitivity and Pareto geometry}

\begin{figure}[t]
  \centering
  \includegraphics[width=\columnwidth]{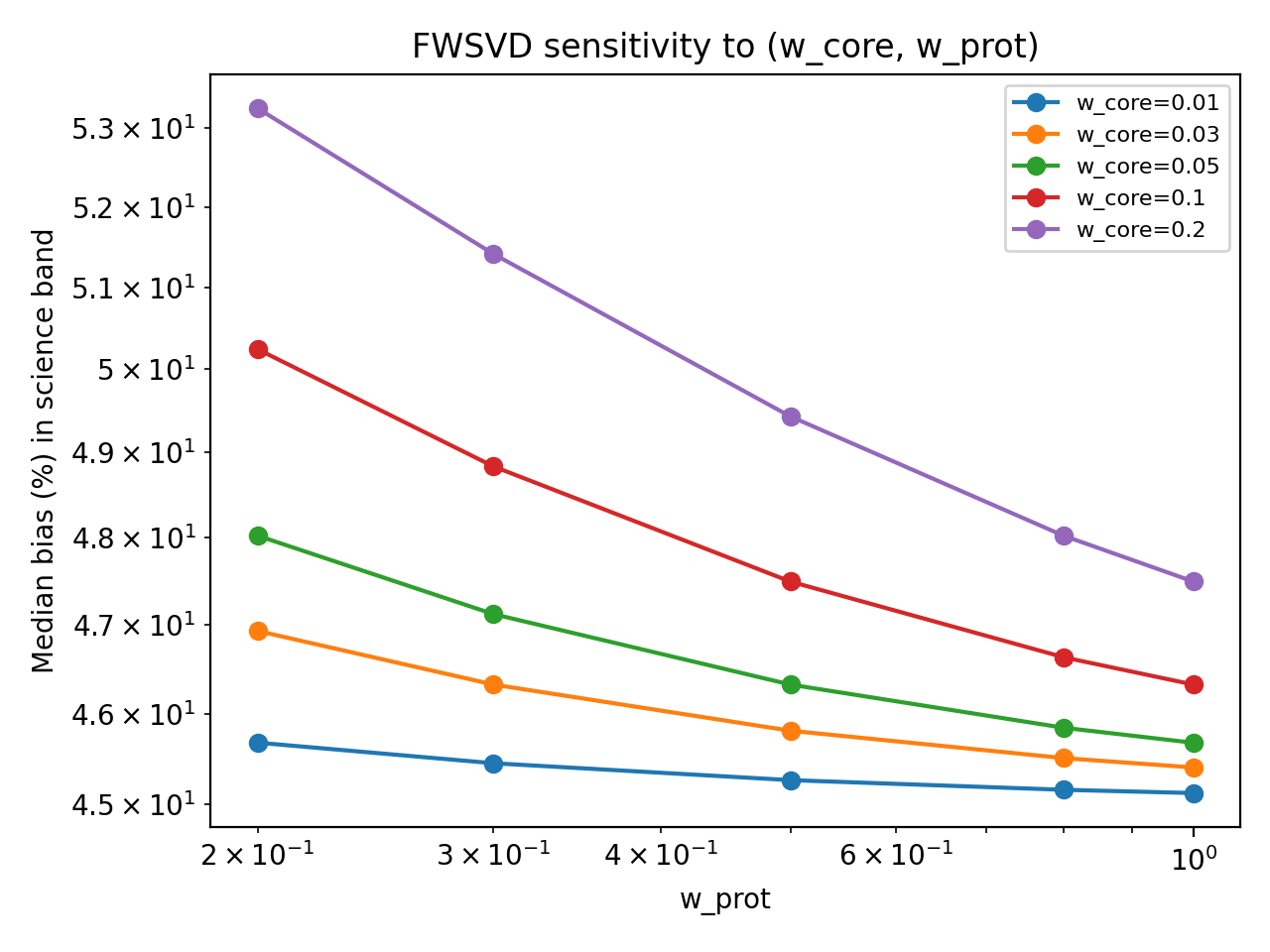}
  \caption{FWSVD sensitivity to $(w_{\rm core}, w_{\rm prot})$. Median relative bias in the science band as a function of $w_{\rm prot}$ for several fixed values of $w_{\rm core}$. Stronger down-weighting of the science core (smaller $w_{\rm core}$) reduces the median bias, but gains saturate once $w_{\rm core} \lesssim 0.03$--$0.05$.}
\label{fig:fwsvd_weights}
\end{figure}

Figure~\ref{fig:fwsvd_weights} shows that for ${\rm SNR}\lesssim1$ in our
synthetic configuration, $w_{\rm core}\simeq0.03$--0.05 and
$w_{\rm prot}\simeq0.3$--0.6 lie close to the minimum median bias across the
grid. Lighter down-weighting rapidly reduces protection of the core, while
more aggressive weights provide only marginal gains but increase sensitivity
to numerical issues. These values therefore define a practical operating
range for FWSVD in this testbed.

\begin{figure*}[t]
    \centering
    \includegraphics[width=\textwidth]{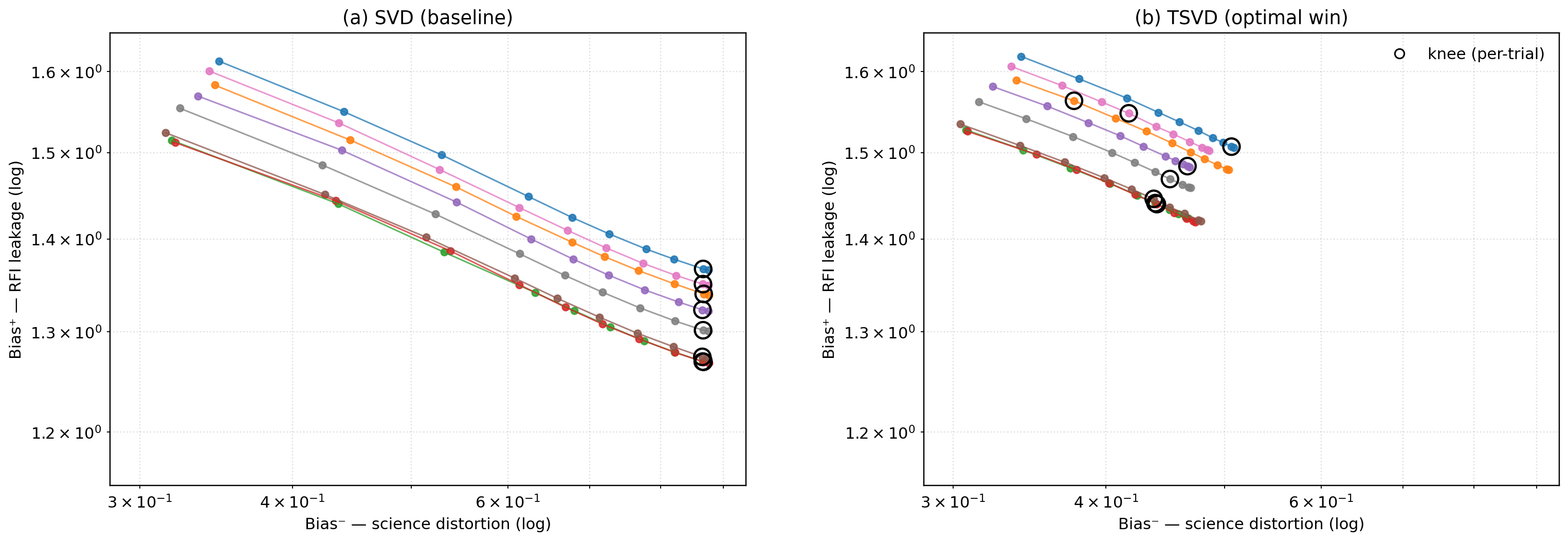}
    \caption{
\textbf{Bias--distortion trade-off in synthetic single-epoch experiments.}
(a) Standard SVD and (b) frequency-weighted SVD (FWSVD) under rank sweep.
Each curve corresponds to one synthetic realization; each point is a rank choice.
The knee of each curve (filled marker) is identified as the point of maximum
curvature in the $\mathrm{Bias}^{+}$--$|\mathrm{Bias}^{-}|$ plane.
In both panels the trade-off curves are shallow rather than L-shaped: no rank
choice simultaneously achieves low residual RFI contamination and low science
distortion. This flatness arises because the available rank budget is insufficient
to isolate the interference subspace---the leading singular modes mix foreground,
RFI, and science-band structure, so each additional rank removed carries both
RFI suppression and science loss in roughly equal measure
(Section~\ref{sec:synthetic_results}).
Frequency weighting shifts the Pareto frontier toward lower distortion at matched
contamination levels, but does not eliminate the underlying trade-off.
    }
    \label{fig:bias_tradeoff_panels}
\end{figure*}

The Pareto projection of the same trade-off is shown in Figure~\ref{fig:bias_tradeoff_panels},
where each panel plots $|\mathrm{Bias}^{-}|$ (science distortion) against $\mathrm{Bias}^{+}$ (residual contamination) for each trial.
The knee of the Pareto front---identified as the point of maximum curvature on the convex hull of the trial cloud, marked by open circles---separates a regime in which increasing rank primarily reduces contamination (below the knee) from one in which it primarily increases distortion (above the knee).
The knee is not sharp because the underlying singular-value spectrum decays gradually rather than exhibiting a clear gap, so that successive rank increments produce diminishing marginal returns in contamination reduction.
This gradual transition is a direct consequence of the mixed-subspace structure: when science and RFI energy are distributed across multiple singular modes with comparable amplitudes, no single rank increment can decisively separate the two components.

\section{Application to HERA Data}\label{sec:hera}

We now apply the diagnostics developed on the synthetic testbed to
real single-epoch HERA snapshots. Our goals are (i) to test whether
the \emph{mixed singular-subspace} structure, the contamination--distortion
trade-off, and the resulting \emph{error floor} observed in controlled
experiments also appear in real data, and (ii) to define a minimal QA
procedure that can be deployed in operational pipelines. As operational
proxies we use (1) the \emph{residual power in heavily flagged channels}
and (2) deviations from \emph{spectral smoothness} within the protected
science band. In the HERA application, the residual-contamination proxy is operationalized as residual power in heavily flagged channels.

\subsection{Data Products and Construction}

We use single-epoch dynamic spectra from the 2022 April 3 HERA observations,
drawn from the publicly available dataset of \citet{Mesarcik2022},
archived at Zenodo (DOI: \href{https://doi.org/10.5281/zenodo.6724065}{10.5281/zenodo.6724065}).
We analyze $N_{\rm snap}=420$ snapshots from the first array, each with
$(T_{\rm raw},F)=(512,512)$, and use a trimmed subset of $T=420$ time samples
after removing edge segments with elevated flagging \citep{DeBoer2017}.
We fix a single interferometric baseline (index~0) and the stored polarization (XX).
A summary of the HERA data products and preprocessing is provided in Table~\ref{tab:hera_summary}.

We collapse the calibrated complex visibility $V(t,\nu)$ into its amplitude
$|V(t,\nu)|$, which serves as a real-valued proxy for the power in each
time--frequency pixel. The final analysis matrix $D\in\mathbb{R}^{T\times F}$
is constructed after median--MAD normalization.
The snapshots cover the 50--225\,MHz band with uniform channel spacing
$\Delta\nu \approx 0.3418\,\mathrm{MHz}$ and time resolution $\Delta t=1.0\,\mathrm{s}$.
For the HERA analysis, we define the science core as 140--160\,MHz, the science
band as 130--170\,MHz, and the outside band as the remaining channels.
The synthetic channel spacing ($\Delta\nu=0.05$\,MHz) is finer than the HERA spacing
($\approx 0.34$\,MHz) in order to resolve narrowband RFI more clearly; this affects
the apparent rank of narrowband structure but not the underlying non-identifiability
mechanism, which is driven by subspace overlap.

\subsection{Evidence for mixed modes}\label{subsec:mixed_modes_hera}

Our first diagnostic is a mixed-mode inspection. The leading singular
modes exhibit co-located foreground/instrument-like smooth structure
and narrowband or comb-like RFI signatures within the \emph{same}
modes (Figure~\ref{fig:hera_pareto}a).
This is the real-data analogue of the synthetic mixed-mode structure shown in
Figure~\ref{fig:singular_mode_mixing}b: in both cases the dominant singular
mode captures a superposition of smooth and narrowband components.
This implies that the HERA snapshots occupy the same mixed-subspace regime as the synthetic
testbed: low-rank projection that removes the leading modes necessarily
suppresses RFI at the cost of \emph{unavoidable} distortion in the
protected science band.

We note that the synthetic testbed models only narrowband comb-like RFI, whereas real HERA data
also contain broadband RFI (e.g., from digital electronics or wideband transmitters).
Broadband interference that is spectrally smooth will tend to be absorbed into the same
low-rank subspace as the astrophysical foreground, reinforcing rather than alleviating the
mixed-mode structure. If the broadband RFI has temporal structure distinct from the foreground,
it may occupy additional singular modes, but this increases the effective rank of the
contaminant without removing the subspace overlap with science. Our narrowband-only synthetic
model therefore represents a conservative scenario; the inclusion of broadband RFI would
generally strengthen the non-identifiability conclusion.

\subsection{Rank sweeps, proxy metrics, and Pareto views}\label{subsec:hera_rank_pareto}

\begin{figure*}[t]
    \centering
    \includegraphics[width=0.95\textwidth]{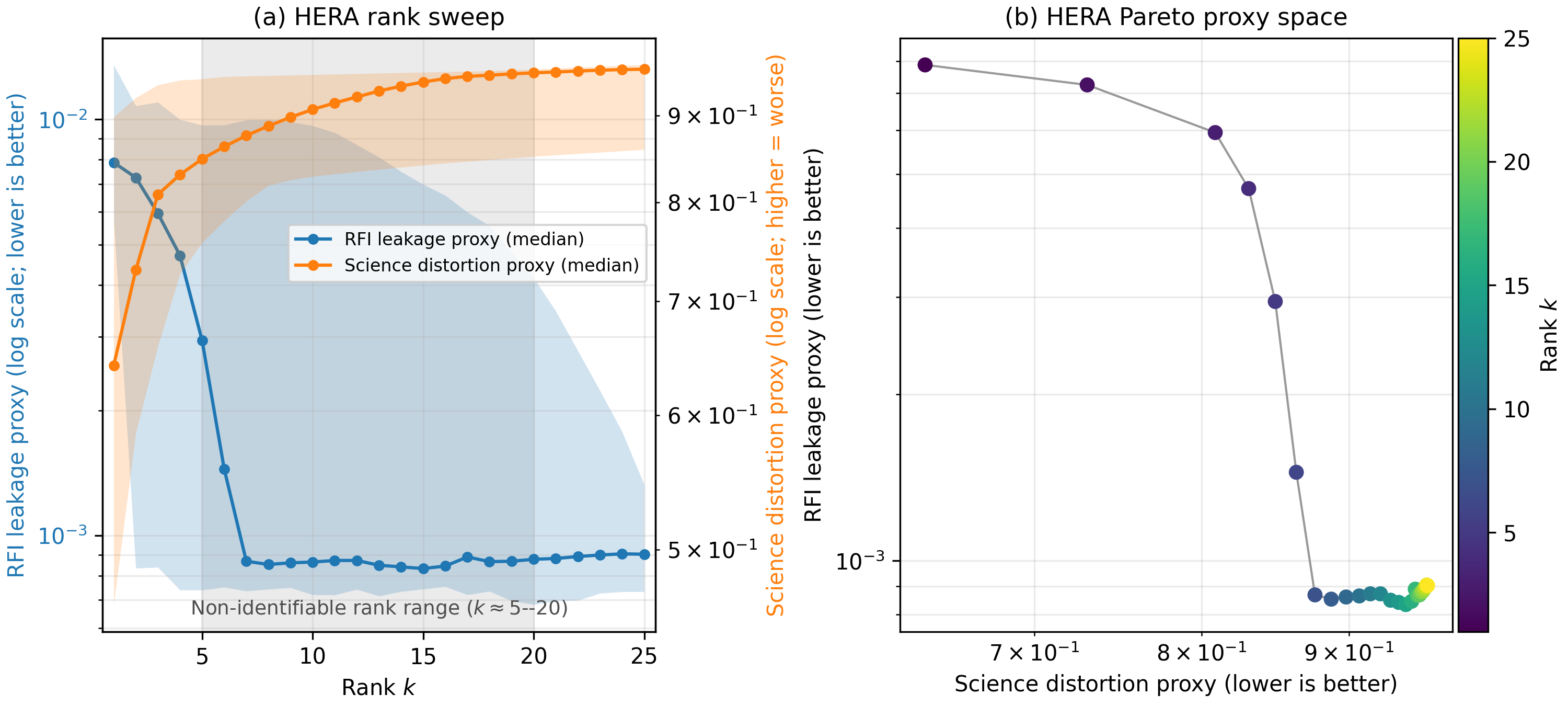}
    \caption{
    \textbf{HERA single-epoch rank sweep and Pareto proxy diagnostics.}
     (a) Residual-contamination proxy (blue, lower is better) decreases with increasing rank while science-distortion proxy (orange, lower is better) increases. The shaded region indicates the non-identifiable rank range (approximately $k=5$--$20$) where both errors remain high. (b) Pareto projection of the same proxies (colored by rank). The front exhibits a clear knee and non-zero error floor, showing that no single rank simultaneously minimizes residual contamination and distortion. Compare with the synthetic Pareto panels in Figure~\ref{fig:bias_tradeoff_panels}.
    }
\label{fig:hera_pareto}
\end{figure*}

We perform rank sweeps over $k$ and summarize the contamination--distortion
trade-off in a Pareto view using the two proxy metrics above.
For calibration, we additionally construct pseudo-snapshots by
injecting a weak EoR-like spectral feature into a subset of snapshots;
in this case the injected science spectrum is known and the recovery
error can be measured directly as a function of rank.
For both TempSVD- and FWSVD-type variants, the science-core RMSE decreases
rapidly from $k=1$ to $k\simeq 3$--5 and then saturates, revealing an
\emph{error floor} beyond which increasing rank yields little or no
improvement. Frequency weighting shifts the operating point toward a
relatively safer region and improves robustness, but does not remove
the floor. For real snapshots the true sky spectrum is unknown, so an
absolute bias cannot be computed; instead, the rank dependence of the
proxies and the Pareto frontier (Figure~\ref{fig:hera_pareto})
demonstrate the same non-identifiability geometry. In particular, the
HERA Pareto envelope shows a knee and does not approach the origin,
indicating that single-epoch low-rank cleaning is \emph{structurally}
non-identifiable rather than merely under-tuned.
The Pareto projection for HERA (Figure~\ref{fig:hera_pareto}b) should be compared
directly with the synthetic Pareto panels in Figure~\ref{fig:bias_tradeoff_panels}:
both exhibit the same qualitative structure---a non-zero error floor and a knee that
prevents the operating point from reaching the origin---confirming that the
non-identifiability is not an artifact of the synthetic setup.

\subsection{Limits of proxy metrics}\label{subsec:hera_proxy_limits}

\begin{figure}[t]
  \centering
  \includegraphics[width=\columnwidth]{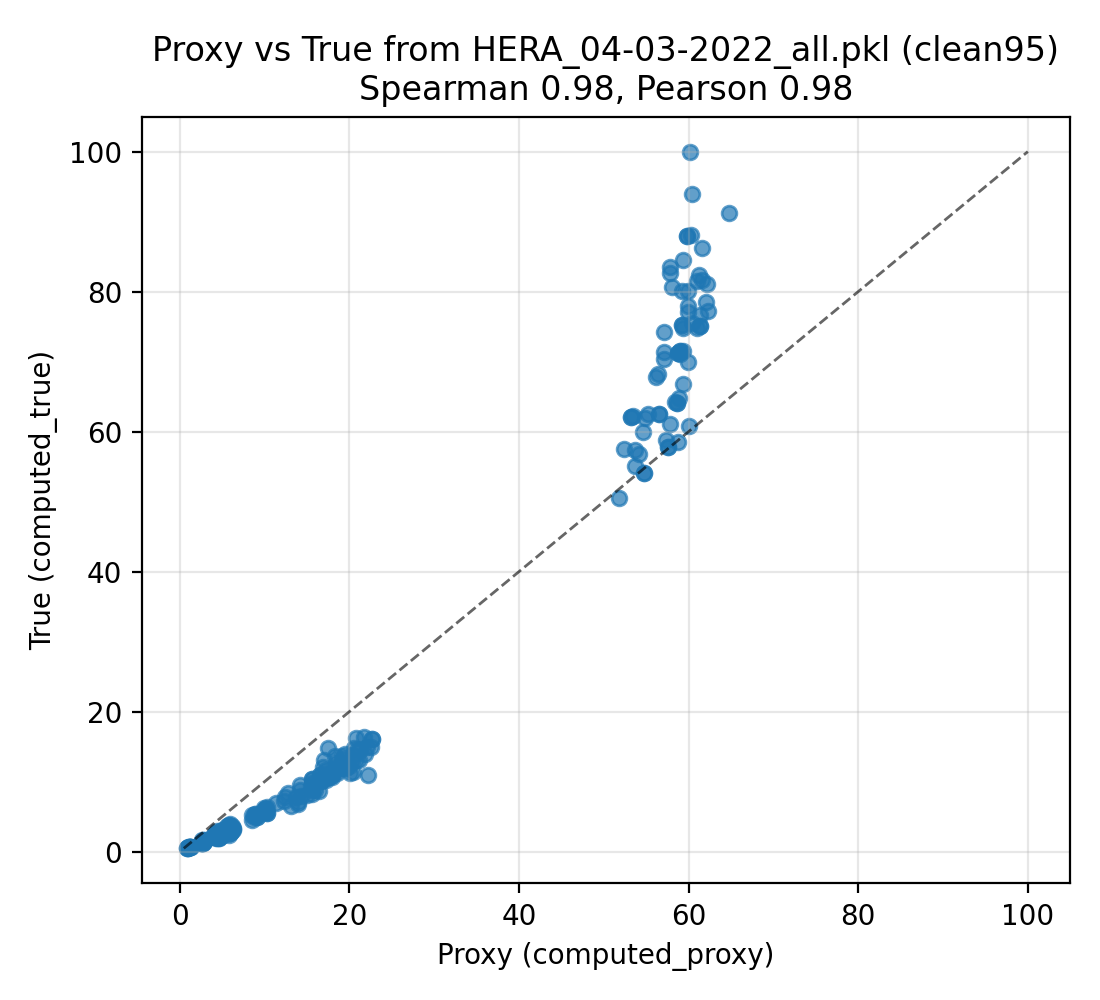}
  \caption{Validation of the residual-based proxy metric on HERA injection tests. The reported correlations are conditional on restricting to the unsaturated regime (excluding trials with $\mathrm{proxy}>95$ and applying the same detection/thresholding rules used in the analysis). Within this unsaturated dynamic range the proxy tracks the true injection--recovery error closely (Spearman $\rho=0.98$, Pearson $r=0.98$), but saturation and threshold effects can break snapshot-level calibration.}
  \label{fig:proxy_correlation}
\end{figure}

In the HERA injection tests we can directly compare the proxy metrics
to the true injection-recovery error. While a rank-level average trend
is present, the correlation weakens at the level of individual
snapshots (Figure~\ref{fig:proxy_correlation}). Even when the Pearson
correlation is large on average (e.g., $\gtrsim 0.8$ across many
snapshots), there exists a tail of cases in which proxy values appear
benign while science distortion is substantial. Proxy metrics should
therefore be interpreted not as per-snapshot estimators of true error,
but as \emph{operational risk indicators} used to summarize rank trends
and to define QA envelopes.

Mathematically, this behavior can be traced to the normalization and masking
structure of the proxy metrics.
Recall that the science-distortion metric ultimately derives from the
channel-wise bias $B(\nu)$ defined in Equation~\eqref{eq:relative-bias}
and the integrated absolute error $\mathrm{IAE}_{\mathrm{core}}$ in
Equation~\eqref{eq:iae-core}.
In heavily flagged intervals, AOFlagger and subsequent masking effectively
remove the most contaminated channels from both the numerator and the
denominator of these expressions, while the remaining channels are
renormalized by a fixed dynamic-range scale (e.g., a MAD-based estimator)
and the stabilizing constant $\epsilon$ defined in Equation~\eqref{eq:relative-bias}.
When the true science spectrum $S(\nu)$ is very small in the surviving
channels, the denominator $\max(|S(\nu)|,\epsilon)$ saturates at $\epsilon$
and the proxy becomes insensitive to additional suppression of $S(\nu)$,
even though the absolute error $|\hat{S}(\nu)-S(\nu)|$ remains large.
Conversely, when bright residual RFI contamination is confined to a small number of heavily
masked channels, the residual-power proxy can be driven to deceptively low values.
These effects explain why rank-averaged diagnostics can exhibit a tight
correlation between proxy and injected error, while snapshot-level
realizations develop a long tail of under-estimated failures.

\subsection{Minimum QA procedure}\label{subsec:hera_minqa}

The HERA results motivate the following minimum QA procedure for any
single-epoch low-rank cleaning configuration:
\begin{enumerate}
\item \emph{Mixed-mode check:} inspect the spectral structure of the
leading singular vectors to determine whether protected science
features plausibly contaminate the modes slated for subtraction.
\item \emph{Rank sweep \& Pareto view:} run rank sweeps on a small set of
representative snapshots and plot contamination--distortion proxies in a
Pareto diagram to identify the presence of an error floor and the
relative ``safer'' rank region.
\item \emph{Proxy calibration:} calibrate chosen proxy metrics against
bias/failure diagnostics on a validation subset; then monitor the
proxies as low-cost QA signals in large-scale processing.
\end{enumerate}
In summary, frequency weighting can be a practical tuning knob that
moves the operating point along the Pareto front, but because a
non-zero error floor persists it should not be presented as a solution
to the single-epoch non-identifiability. Operational pipelines should
therefore document the adopted weights, the induced Pareto shift, and
the parameter ranges that satisfy the QA envelope.

\begin{figure*}[t]
  \centering
  \includegraphics[width=0.95\textwidth]{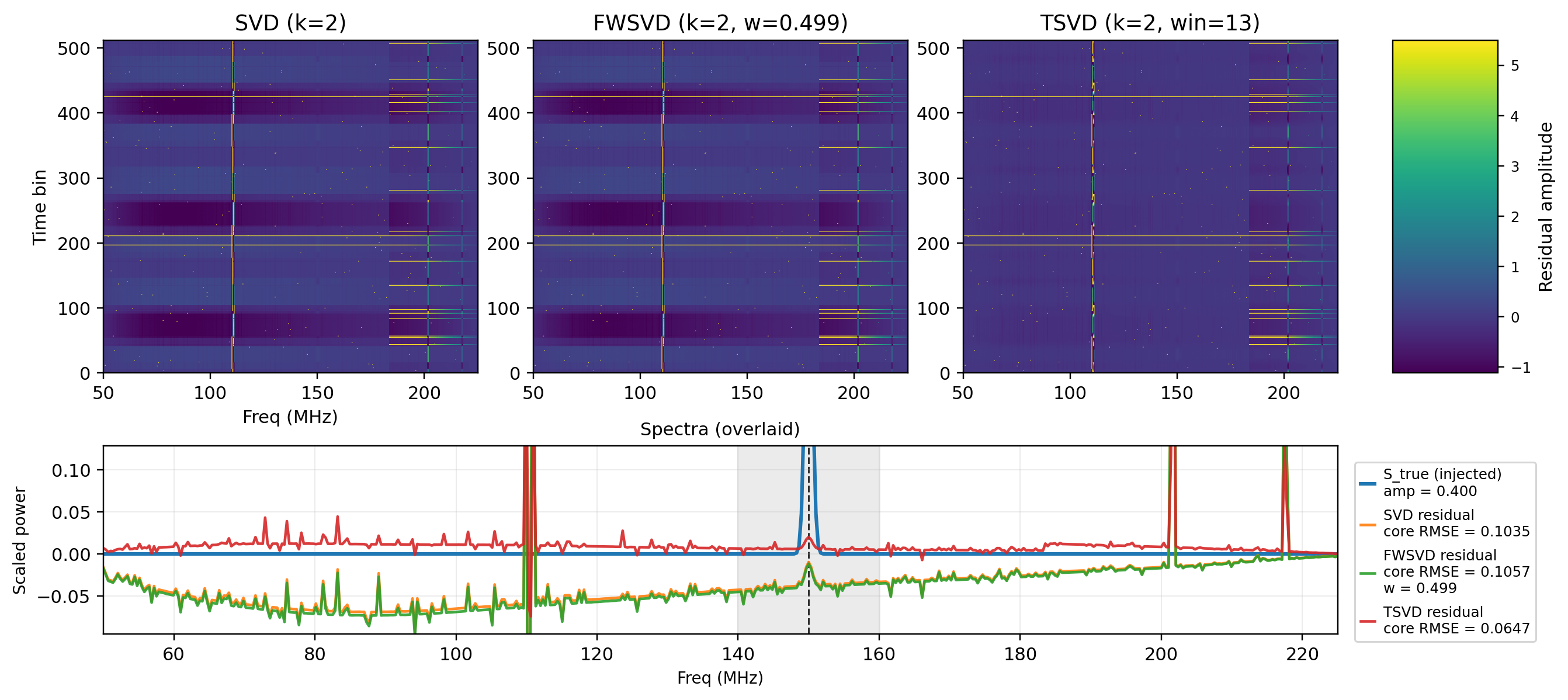}
  \caption{\textbf{Rank-2 truncation failure on HERA data.}
Residual structure after rank-$k=2$ low-rank truncation on a single-epoch
HERA dynamic spectrum with an injected EoR-like signal at 150\,MHz
(amplitude $=0.4$ in MAD-normalized units; see Appendix~\ref{app:eor_injection}
for the injection model and amplitude choice).
\textit{Top:} Time--frequency residual matrices $D - D_k$ for SVD, FWSVD,
and TempSVD. While all methods remove the dominant low-rank foreground modes,
each leaves distinct method-dependent artifacts: standard SVD retains
broadband foreground residuals; FWSVD redistributes variance outside the
weighted core; and TempSVD introduces horizontal smoothing features associated
with temporal averaging.
The dark vertical stripes visible in the SVD and FWSVD panels (e.g., near
time bins $\sim$75, 250, and 410) correspond to short intervals of elevated
RFI activity or instrumental instability that are not fully captured by the
rank-2 model; these stripes are absent in the TempSVD panel because the
temporal smoothing kernel averages over such transient features, effectively
spreading their energy across neighboring time bins.
\textit{Bottom:} Frequency-averaged residual spectra overlaid with the
injected signal (blue). Vertical shading marks the protected EoR window
(140--160\,MHz), with the dashed line indicating the injection frequency.
Despite comparable core RMSE values (SVD: 0.10, FWSVD: 0.11, TempSVD: 0.06),
none of the methods cleanly isolates the injected signal from foreground
residuals. The qualitative differences in distortion patterns, without a
clear ordering in performance, illustrate the operational non-identifiability
of single-epoch low-rank foreground cleaning.
}
\label{fig:hera_rank1_failure}
\end{figure*}

Figure~\ref{fig:hera_rank1_failure} illustrates the method-dependent residual
structure after rank-2 truncation on a representative HERA snapshot with an
injected EoR-like signal.

\section{Conclusion and outlook}
\label{sec:conclusion}

Using controlled synthetic experiments and HERA single-epoch data, we have
shown that low-rank (SVD/PCA-family) RFI mitigation in the single-epoch regime
faces an operational non-identifiability: when the scientific component and
structured RFI share the same singular subspace, no rank choice can
simultaneously suppress residual RFI contamination and preserve the science signal.
This manifests as mixed singular modes, rank sweeps with monotonic under/over
cleaning trade-offs, and Pareto fronts with non-zero bias floors.
Frequency-weighted SVD and related variants can shift the operational
trade-off---moving the Pareto knee to safer regions and reducing sensitivity to
small fluctuations---but do not eliminate the bias floor when subspace overlap
is strong. Applying the same diagnostics to HERA single-epoch snapshots
reveals the same structure as in the synthetic testbed, bridging the gap
between controlled failure-mode analysis and real data.

Our analysis is intentionally conservative. We focus on a single-epoch,
single-baseline, single-polarization configuration; any diversity across time,
baseline length, or polarization is treated as external to the problem rather
than as an additional axis for identifiability. In a full EoR analysis this
diversity can and should be exploited, and our results should be interpreted as
a lower bound on what is achievable without it. The synthetic testbed is
anchored on a harmonic-comb RFI model motivated by recent satellite work
\citep[e.g.,][]{DiVruno2023a, Bassa2024} and on a
single, time-invariant spectral feature; broadband, drifting, or bursty RFI
scenarios may introduce additional structure but do not remove the possibility
of mixed singular directions. In the HERA application the true science
signal is unknown, so we rely on proxy metrics (smoothness, residual structure
in heavily flagged regions) rather than direct bias measurements, and we explicitly demonstrate that such proxies cannot be interpreted as realization-level error estimators.

These caveats point to future work: (i) quantifying how much multi-epoch and
multi-baseline diversity is needed to move Pareto fronts closer to the origin;
(ii) combining low-rank decompositions with physical priors or non-linear
models of the foregrounds, beam, or instrument; and (iii) extending the
synthetic testbed to a broader class of RFI morphologies and to other
low-frequency arrays.

A simple operational rule emerges:
\begin{enumerate}
\item Inspect the leading singular modes: if they clearly mix smooth
foreground-like structure and narrowband or comb-like RFI, treat the snapshot
as intrinsically non-identifiable \emph{a priori}.
\item Run a short rank sweep and check whether both a science-core bias metric
and a residual-contamination proxy fall below pre-defined QA thresholds.
\item If no such rank exists, avoid aggressive low-rank subtraction and rely
instead on conservative masking and on multi-epoch or multi-baseline diversity
for scientific interpretation.
\end{enumerate}
In this view, single-epoch low-rank cleaning is a risk-managed tool with an
explicit QA envelope, not a universally safe default, complementing existing
protection criteria and RFI-mitigation practice in radio astronomy
\citep[e.g.,][]{ITUR2005RA769, FridmanBaan2001, Offringa2010, Offringa2012}.

\section*{Acknowledgments}

The author thanks colleagues and community discussions on RFI impacts and mitigation practices that motivated this diagnostic framing.
This work makes use of data from the Hydrogen Epoch of Reionization Array
(HERA; \url{https://reionization.org/}), which is supported by the National
Science Foundation and the Gordon and Betty Moore Foundation. We gratefully
acknowledge the HERA collaboration for making their data products accessible.
HERA is hosted by the South African Radio Astronomy Observatory, which is a
facility of the National Research Foundation, an agency of the Department of
Science and Innovation.
This work builds upon exploratory experiments conducted in the public code repository \url{https://github.com/kimeujin03-droid/single-epoch-rfi-mitigation}.

\appendix
\section{Synthetic data parameters and signal-generation recipe}
\label{app:synthetic}

Table~\ref{tab:synthetic-params} lists the fixed synthetic-data parameters
used throughout the controlled toy experiments, unless stated otherwise.
The configuration matches the simulation settings used to generate all
synthetic figures in this work.
Our synthetic testbed follows standard practice in time--frequency modelling
by combining smooth astrophysical components with narrowband and comb-like
RFI structures \citep[e.g.,][]{Leshem2000}, while allowing full control over
signal-to-interference overlap and relative amplitudes.
The signal-generation recipe is:

\begin{deluxetable}{ll}
  \tablecaption{
Synthetic dynamic-spectrum parameters used in the controlled comb+line
experiments (Figures~2--4). Values match the configuration used in the SVD diagnostics setup.
The listed FWSVD weights are the nominal values used to render Figures~2--4; Section~\ref{sec:synthetic_results} explores a wider weight grid and reports a practical operating range.
\label{tab:synthetic-params}}
 \tablehead{
\colhead{Quantity} & \colhead{Value}
}\startdata
Time resolution $\Delta t$          & $1.0~\mathrm{s}$ \\
Total duration                      & $60~\mathrm{s}$ ($T = 60$ samples) \\
Frequency range                     & $0$--$12~\mathrm{MHz}$ \\
Channel width $\Delta\nu$           & $50~\mathrm{kHz}$ ($F = 240$ channels) \\
Science feature                     & Time-invariant Gaussian line in frequency \\
Science line center                 & $6.0~\mathrm{MHz}$ \\
Science line width                  & $\sigma_\nu = 0.2~\mathrm{MHz}$ \\
Science amplitudes                  & Broad pedestal $0.03$, line peak $0.10$ (arb.\ units) \\
Science band (evaluation window)    & $5.5$--$6.5~\mathrm{MHz}$ \\
Protected core band                 & $5.8$--$6.2~\mathrm{MHz}$ \\
Comb interference lines             & 5 lines at 5.6, 5.8, 6.0, 6.2, 6.4 MHz \\
Comb line width                     & $\sigma_\nu = 0.02~\mathrm{MHz}$ \\
Comb peak amplitude                 & $10.0$ (grid: 5, 10, 15, 20) \\
Time-localized burst envelope       & Gaussian in time, center $5~\mathrm{s}$, width $1~\mathrm{s}$ \\
Broadband spectral slope            & $0.03$ across band \\
Sinusoidal ripple amplitude         & $0.03$ \\
Sinusoidal ripple period            & $1.5~\mathrm{MHz}$ \\
Thermal noise $\sigma$              & $0.001$, i.i.d.\ Gaussian \\
Default truncation rank $k$         & $k = 1$ (sweep: $k=1$--$50$) \\
Example FWSVD weights (Figs.~2--4)  & $w_{\rm core}=0.1$, $w_{\rm prot}=0.3$ \\
\enddata
\end{deluxetable}

\begin{table*}[t]
\centering
\caption{HERA single-epoch data product and preprocessing summary.}
\label{tab:hera_summary}
\begin{tabular}{lll}
\toprule
Item & Value & Notes \\
\midrule
Instrument / date & HERA, 2022-04-03 & single-epoch snapshots \\
Data source & \citet{Mesarcik2022}, Zenodo & DOI: 10.5281/zenodo.6724065 \\
Observation window & $\sim$19:42--20:42 UTC & from Zenodo metadata \\
Data format & Python pickle (list of 4 arrays) & loaded via \texttt{pd.read\_pickle} \\
Array 0 shape & $(420, 512, 512, 1)$ float16 & visibility amplitude \\
Array 1 shape & $(420, 512, 512, 1)$ bool & AOFlagger mask \\
Number of snapshots $N_{\rm snap}$ & 420 & first axis of array 0 \\
Per-snapshot matrix size & $T=420$, $F=512$ & time samples $\times$ frequency channels \\
Frequency span & 50--225\,MHz & stated analysis band \\
Channel spacing $\Delta\nu$ & $0.341796875\,\mathrm{MHz}$ & measured from frequency axis \\
Time resolution $\Delta t$ & $1.0\,\mathrm{s}$ & verified from metadata/header \\
Snapshot duration $T_{\rm snap}$ & $420\,\mathrm{s}$ & $T_{\rm snap}=T\times\Delta t$ \\
Baseline & Baseline index 0 & fixed per snapshot \\
Polarization & XX & fixed \\
Input quantity & $|V(t,\nu)|$ & amplitude of calibrated visibility \\
Normalization & median--MAD & applied to form $D\in\mathbb{R}^{T\times F}$ \\
RFI mask & AOFlagger boolean mask & separate array (not NaNs) \\
Flagged fraction & mean 2.75\%, median 2.24\% & 10/90\%: 1.13/5.88\% \\
Science core (protected) & 140--160\,MHz & redshifted 21\,cm at $z\approx 7.9$--$9.1$ \\
Science band (evaluation) & 130--170\,MHz & core $\pm 10$\,MHz \\
Injection example & 150\,MHz, amp = 0.4, $\sigma_\nu=0.5$\,MHz & rank-2 proxy validation \\
\bottomrule
\end{tabular}
\end{table*}

\section{EoR-like injection model for HERA proxy validation}
\label{app:eor_injection}

For the HERA injection--recovery tests (Figures~\ref{fig:proxy_correlation}
and~\ref{fig:hera_rank1_failure}), we inject a narrowband Gaussian spectral
feature centered at 150\,MHz with a width $\sigma_\nu = 0.5$\,MHz and a peak
amplitude of 0.4 in MAD-normalized units (after rescaling the EoR window to
unit RMS). The feature is constant in time
and added to the amplitude matrix $|V(t,\nu)|$ after normalization.

This injection is intentionally simple: it is designed as a controlled probe of
the recovery fidelity of single-epoch low-rank cleaning, not as a physically
realistic model of the 21\,cm signal. The true EoR signal is broadband
($\Delta\nu \gtrsim 10$\,MHz) and spectrally complex
\citep[e.g.,][]{FurlanettoOhBriggs2006, Morales2010};
however, a localized spectral probe provides a sharper, unambiguous recovery
metric than a broadband signal whose distortion could be confused with residual
foreground structure.
Crucially, a narrowband injection represents an \emph{easier} recovery target:
if single-epoch low-rank cleaning distorts even this simple feature, it will
necessarily also distort the broader, weaker true EoR signal.

The amplitude of 0.4 (in EoR-window-RMS-normalized units) was chosen to place
the probe at a signal-to-residual ratio of order unity---comparable to the
residual foreground fluctuations in the science band after rank-2 subtraction---so
that the injection is neither trivially recoverable nor completely buried.
This regime is the most informative for diagnosing non-identifiability:
at much higher amplitude the probe would survive any cleaning method, while at
much lower amplitude detection failure would be attributable to noise rather
than to subspace mixing.
We did not explore a grid of injection amplitudes in this work; varying the
amplitude would shift the detection failure rate $p_{\mathrm{fail}}$ but would
not change the qualitative structure of the Pareto front or the presence of the
bias floor, which are the primary conclusions.

The protected EoR window of 140--160\,MHz corresponds to the redshifted 21\,cm
line at $z \approx 7.9$--$9.1$, consistent with the frequency range commonly
targeted by HERA foreground-avoidance analyses
\citep[e.g.,][]{DeBoer2017, Abdurashidova2022}.

\bibliographystyle{aasjournal}
\bibliography{references}

\end{document}